# CUBES: optical design and analysis of the VLT's high-resolution UV spectrograph past final design review

Lawrence Bissell[*a], Walter Seifert[b], Ariadna Calcines Rosario[c], Hans Dekker[d], Gerardo Avila[e], Luca Oggioni[d], Giorgio Pariani[d], Matteo Genoni[d], Roberto Cirami[f], Stefano Covino[d]

[a]STFC - United Kingdom Astronomy Technology Centre (UK ATC), Edinburgh, UK;
[b]Landessternwarte - Zentrum für Astronomie der Universitat Heidelberg, Heidelberg, Germany;
[c]Durham University - Centre for Advanced Instrumentation - Department of Physics, UK;
[d]INAF-Osservatorio Astronomico di Brera, via E. Bianchi 46, 23807 Merate, Italy;
[e]Optical Lab Baader Planetarium GmbH, Zur Sternwarte 4D-82291 Mammendorf Germany;
[f]INAF-Osservatorio Astronomico di Trieste, via G. B. Tiepolo 11, 34131 Trieste, Italy

## ABSTRACT

CUBES (Cassegrain U-Band Efficient Spectrograph) will be the most efficient UV spectrograph on an 8-10m class telescope, exploiting the ESO VLT's higher telescope efficiency relative to the ESO ELT in ground-based UV wavelengths. The instrument provides wavelength coverage over a range of 300-405nm, with high instrumental efficiency (>37%) and a resolving power of >20,000 (HR) and >6,000 (LR). It will be mounted on the Cassegrain focus of the VLT in Paranal. This paper presents the optical design of the system after the final design review. The optical subsystems detailed include: a pair of UV-optimized atmospheric dispersion correctors designed to function out to a zenith angle of 65°; high throughput HR and LR image slicers used to reformat the field-of-view; and a dispersion-compensating three-lens spectrographic camera design.

## 1. INTRODUCTION

The Cassegrain U-Band Efficient Spectrograph (CUBES) is a UV spectrograph designed to be installed at a Cassegrain focus of one of the Very Large Telescope (VLT) unit telescopes. Due to the mirror coating chosen for the Extremely Large Telescope (ELT), the telescope efficiency of the VLT is significantly higher at wavelengths shorter than 400nm. Therefore, despite the 4.8 times smaller telescope diameter, a high-efficiency U-band spectrograph on the VLT would be a world-leading instrument of its class for the European Southern Observatory (ESO) for the coming decades, complementary to the infrared-optimised ELT. CUBES is designed to be a high-efficiency, intermediate spectral resolution, ground-UV instrument, opening this new discovery space for investigation of novel science cases over a broad range of astrophysics [1].

The following sections detail the optical design after the Final Design Review of the instrument. Firstly an overview of the optical design will be given. Then, the design of the front-end optics will be presented, focusing on the optical design of the atmospheric dispersion correctors. There follows the design of the image slicer, and the spectrograph camera optical design.

## 2. OPTICAL DESIGN OVERVIEW

### 2.1 Optical Design Overview

To achieve the high efficiency required by the CUBES science cases, the instrument will be mounted at the Cassegrain focus of the VLT unit telescopes.. The functional diagram of the CUBES optical design is outlined in Figure 1.

[*]lawrence.bissell@stfc.ac.uk

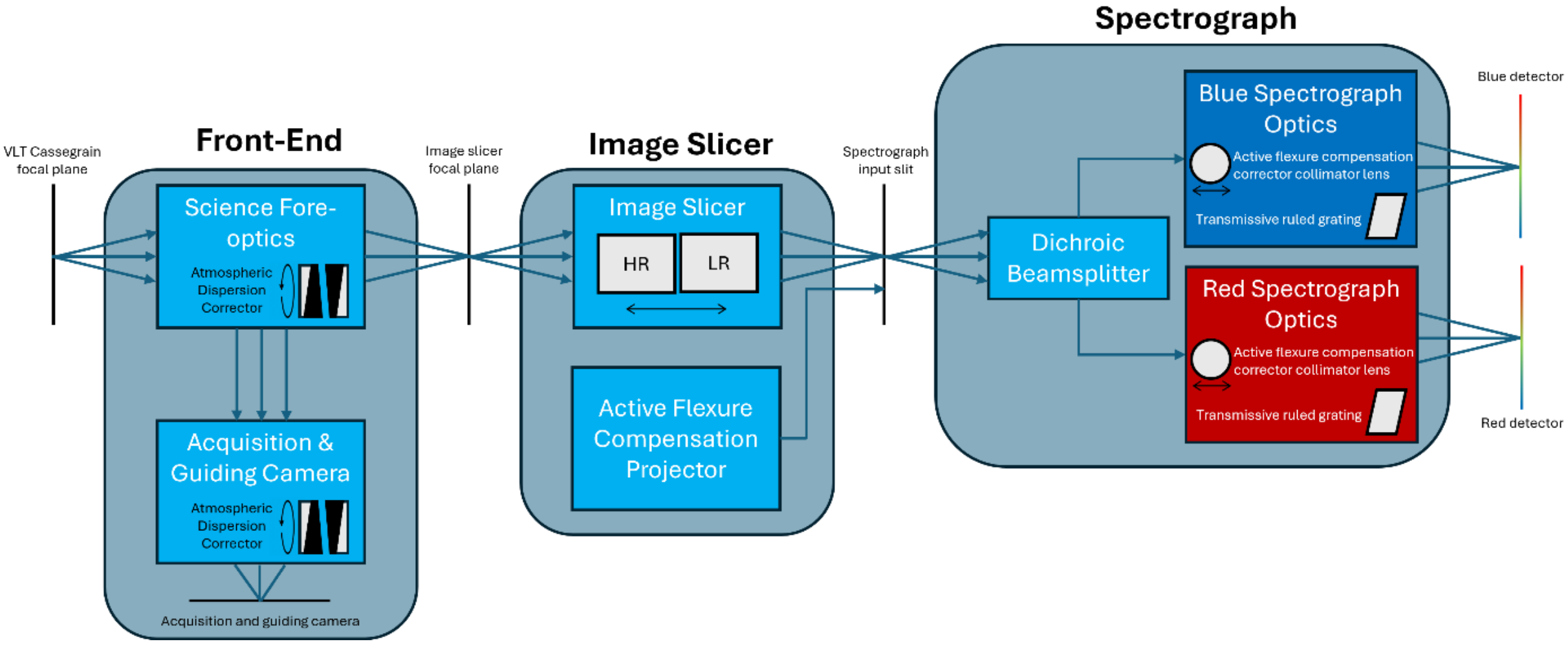

Figure 1. Functional diagram of the CUBES optical design.

Light from the telescope Cassegrain focal plane first enters the front-end optics. This includes:

- a fore-optics sub-system which magnifies the rectangular 1.5” x 10” (HR) or 6” x 10” (LR) science fields-of-view (FOVs) onto the image slicer focal plane and positions the exit pupil at the desired location before the slicer image plane.
- an acquisition and guiding (A&G) sub-system (FOV: 90” x 90”) which is fed by a fold mirror close to the telescope focus, allowing for precise target acquisition and secondary guiding as well as providing an imaging mode to perform photometry.

Both sub-systems also include atmospheric dispersion correctors designed to correct for atmospheric dispersion throughout their operating wavelength ranges.

The image slicers then re-format the rectangular FOVs to generate the spectrograph input slit – decomposing the FOV into six vertically-stacked slices. In addition, light from a Thorium-Argon hollow cathode lamp is projected onto a rectangular slit located ~11mm above spectrograph input slit. This is fed into the spectrograph and periodically read-out on the detector to provide active flexure compensation during science exposures as long as 2 hours.

Light entering the spectrograph is spectrally separated into 2 arms by a dichroic beam splitter. The light from each arm is then collimated and dispersed by a high line-density binary transmission grating. A camera then focuses the resulting spectrum onto the detector (CCD-290-99). The detector has 10µm x 10µm rectangular pixels in a 9216 x 9232-pixel array. The spectrograph camera is designed to provide a spectral sampling of >23µm per input slit width on the detector, i.e., >2.3 pixels/input slit width, to ensure Nyquist sampling.

## 3. FRONT-END OPTICS

### 3.1 Fore-optics

The fore-optics subsystem is designed to transfer and magnify the telescope focal plane onto the image slicer focal plane, at a magnification of 3.79x, to generate an image on the slicer mirrors with a plate-scale of 0.5 arcseconds/mm. To provide this magnification, a doublet lens, composed of Fused Silica and Calcium Fluoride, is placed 304mm after the telescope focal plane to form a collimated beam. This beam then passes through the atmospheric dispersion corrector prisms, which are discussed in detail in 3.3. The light is then folded into the plane of the instrument using a total internal reflection fold prism and then reimaged using another Fused Silica-Calcium Fluoride doublet with a focal length of 1153mm. The reimaging camera doublet is placed in optical contact with the fold prism to prevent ghosts from double reflections between the optical surfaces. Two fold mirrors are then used to transfer the focal plane onto the image slicing mirrors. The pupil is placed 1155mm before the image slicer focus – as desired to ensure optimal performance of the image slicer.

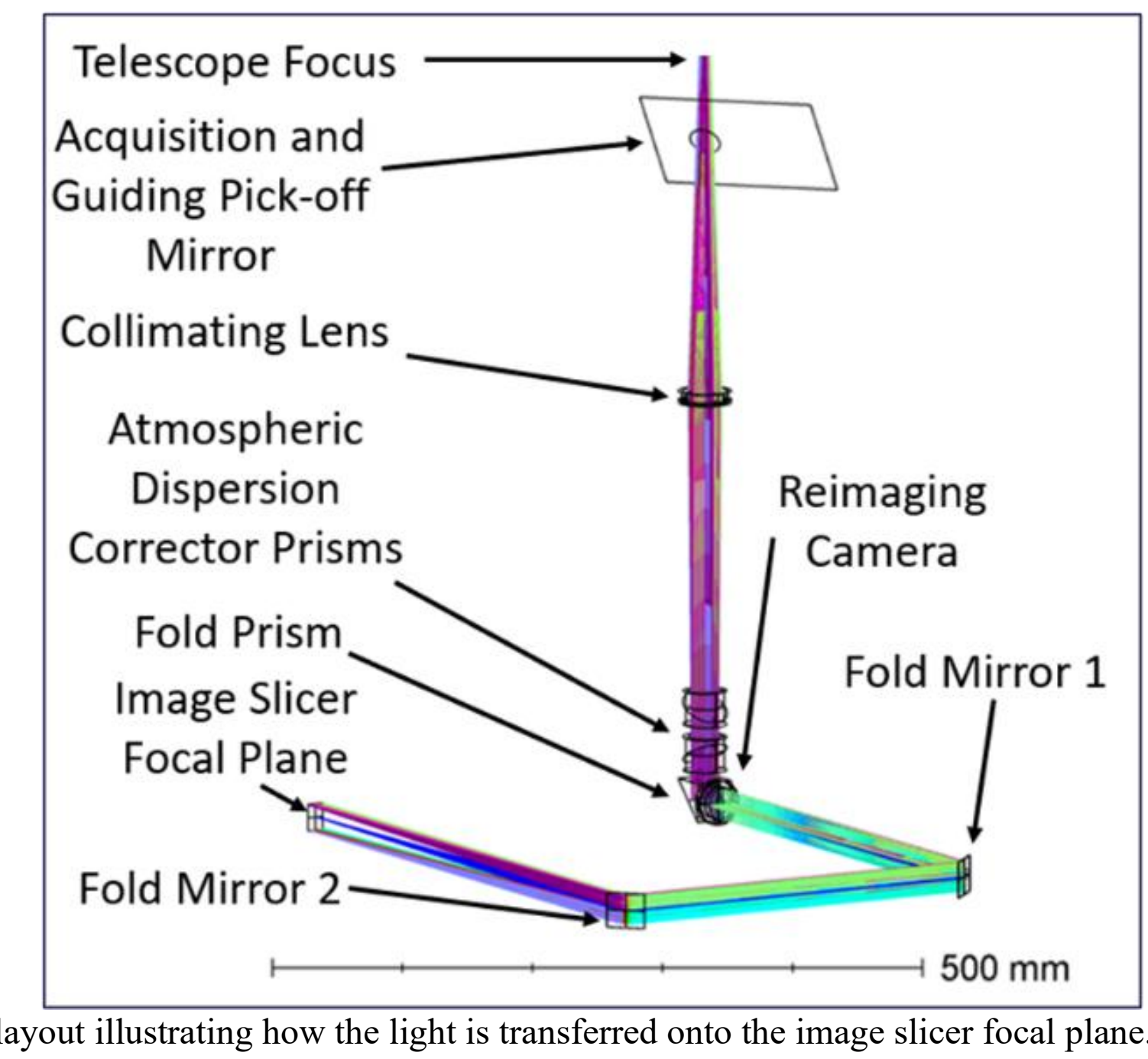

Figure 2. Fore-optics layout illustrating how the light is transferred onto the image slicer focal plane.

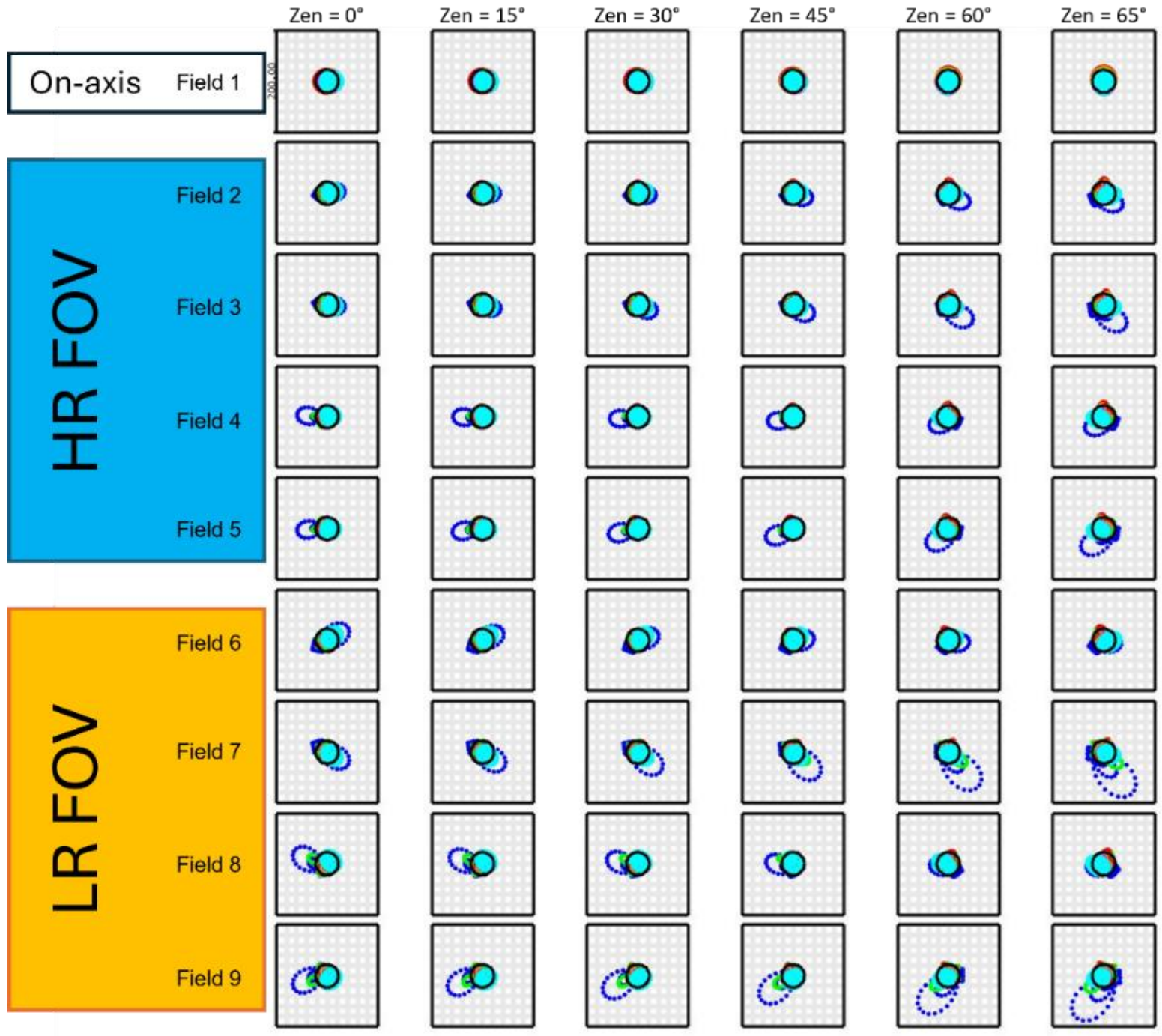

Figure 3. Spot diagrams produced at the slicer focal plane. Columns 1 to 6 correspond to zenith angles 0°, 15°, 30°, 45°, 60°, and 65°. Field 1 corresponds to an on-axis field, fields 2-5 are the corners of the HR image slicer FOV, while fields 6-9 correspond to the corners of the LR image slicer FOV.

The image quality produced by the fore-optics over the 300-405nm wavelength range is shown from the spot diagrams produced at the image slicer focal plane in Figure 3. The scale of each square box is 200µm which is 0.1”, while the RMS diameter of the spots is <0.02”. Given the instrument is seeing-limited this means the image quality of the fore-optics will have a negligible impact on the full-width-half-maximum (FWHM) of astronomical targets at the slicer focal plane.

## 3.2 Acquisition and Guiding Camera

The acquisition and guiding (A&G) camera is positioned after a 45-degree mirror placed 110mm from the telescope focus. This mirror has a hole at its centre to let through the science field to the image slicer. For initial acquisition, the mirror is translated in-plane so also the central part of the field can be viewed. A filter wheel is placed in the optical path allowing for the insertion of filters for the U-band, G-band, and R-band. Science targets can then be acquired using the U-band, while guiding can be performed either in the G-band or R-band depending on the target.

Figure 4 shows the optical layout of the A&G camera. A pair of lenses (Fused Silica and Calcium Fluoride) collimate the 90” x 90” FOV. The collimated beam then passes through the filter and the atmospheric dispersion corrector prisms (see 3.3). A triplet composed of 2 calcium fluoride lenses and a fused silica lens followed by a fused silica singlet lens then reimage the FOV onto the A&G detector (ELSE-I TCCD) focal plane at a plate scale of 7.69 arcseconds/mm.

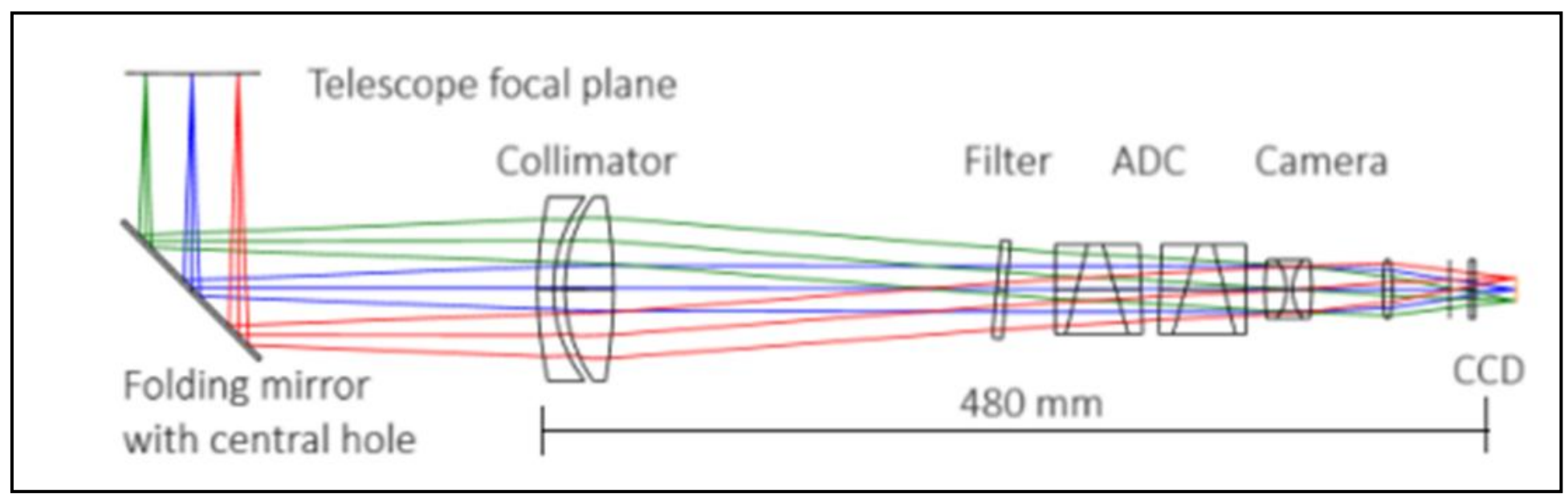

Figure 4. A&G optical layout.

The image quality of the A&G optics is shown in Figure 5 for several fields located around the FOV, over the wavelength range between 300 to 700nm. The scale of each square box is 130µm which corresponds to 1”. The image quality is sufficient (<0.45” RMS diameter) to provide accurate centroiding for both acquisition (U-band) and guiding (G-band/R-band).

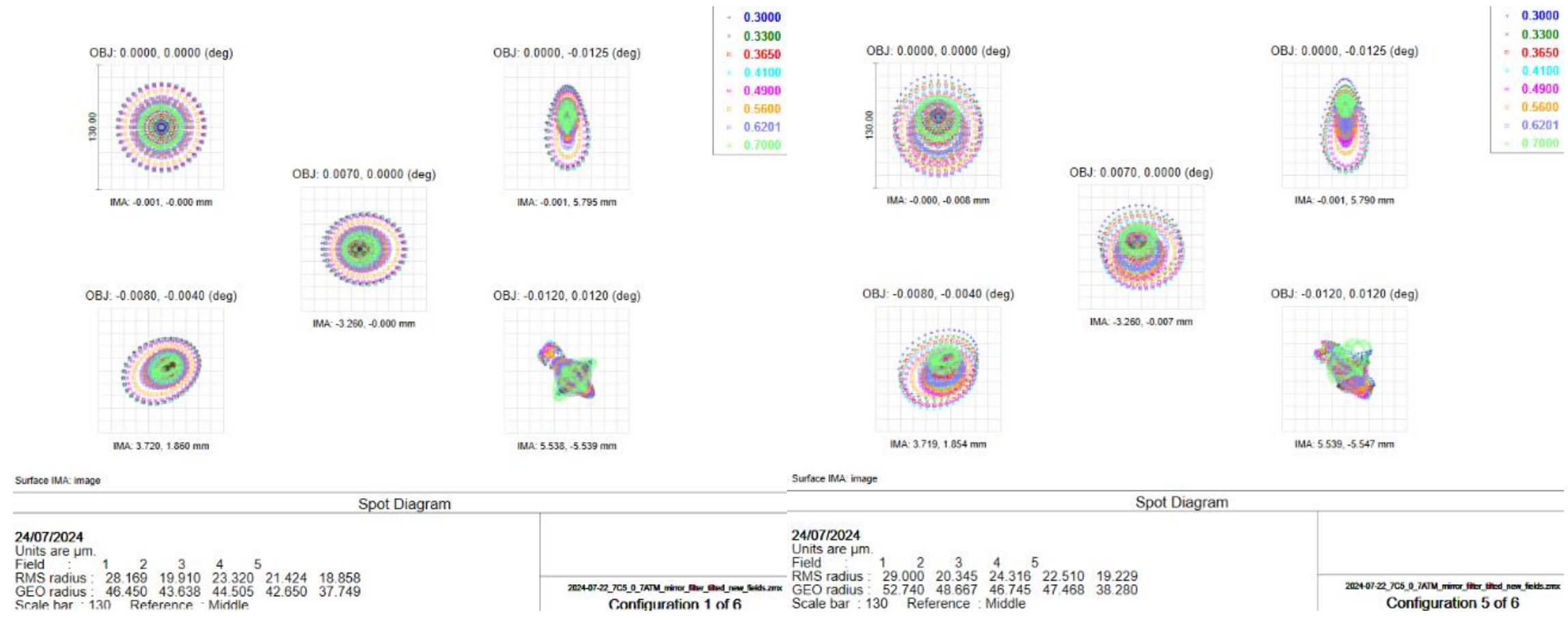

Figure 5. Spot diagram produced at the A&G detector, showing the performance of the A&G optics for zenith angles 0° and 60°.

### 3.3 Atmospheric Dispersion Correction

Atmospheric dispersion is a phenomenon in which light of different wavelengths differentially refracts as it enters the Earth's atmosphere. The effect of atmospheric dispersion is increased the further the telescope is angled from zenith, since light passes through the atmosphere at a steeper angle; and becomes particularly severe in UV.

Over the wavelength ranges of the fore-optics (300-405nm) and the A&G (300-700nm) there is significant atmospheric dispersion:

- Between 300 and 405nm there is 2.45" of differential dispersion at 60° from zenith (3.03" at 65°).
- Between 300 and 700nm there is 4.25" of differential dispersion at 60° from zenith (5.25" at 65°).

Figure 6 shows the atmospheric dispersion expected at the VLT for standard atmospheric conditions.

The fore-optics and A&G sub-systems have been designed to provide atmospheric dispersion correction over their operating wavelength ranges and full FOVs from 0-65° from zenith for the fore-optics and 0-60° from zenith for the A&G sub-system. To provide the atmospheric dispersion correction, the fore-optics use a pair of counter-rotating Amici prisms as the atmospheric dispersion correctors (ADCs), while the A&G uses a pair of counter-rotating double-Amici prisms, as shown in Figure 7. As the angle from zenith is increased, the two prisms rotate to disperse the light in the opposite direction, precisely cancelling the atmospheric effect. To do this, the wedge angles of the prisms were optimised in Zemax OpticStudio, such that when rotated 180° relative to each other, the prisms correct for the differential dispersion at the maximum defined zenith angle. OpticStudio's in-built atmospheric model has been shown to be inaccurate in the blue [2], so the atmosphere was modelled in Zemax using the method proposed in [3].

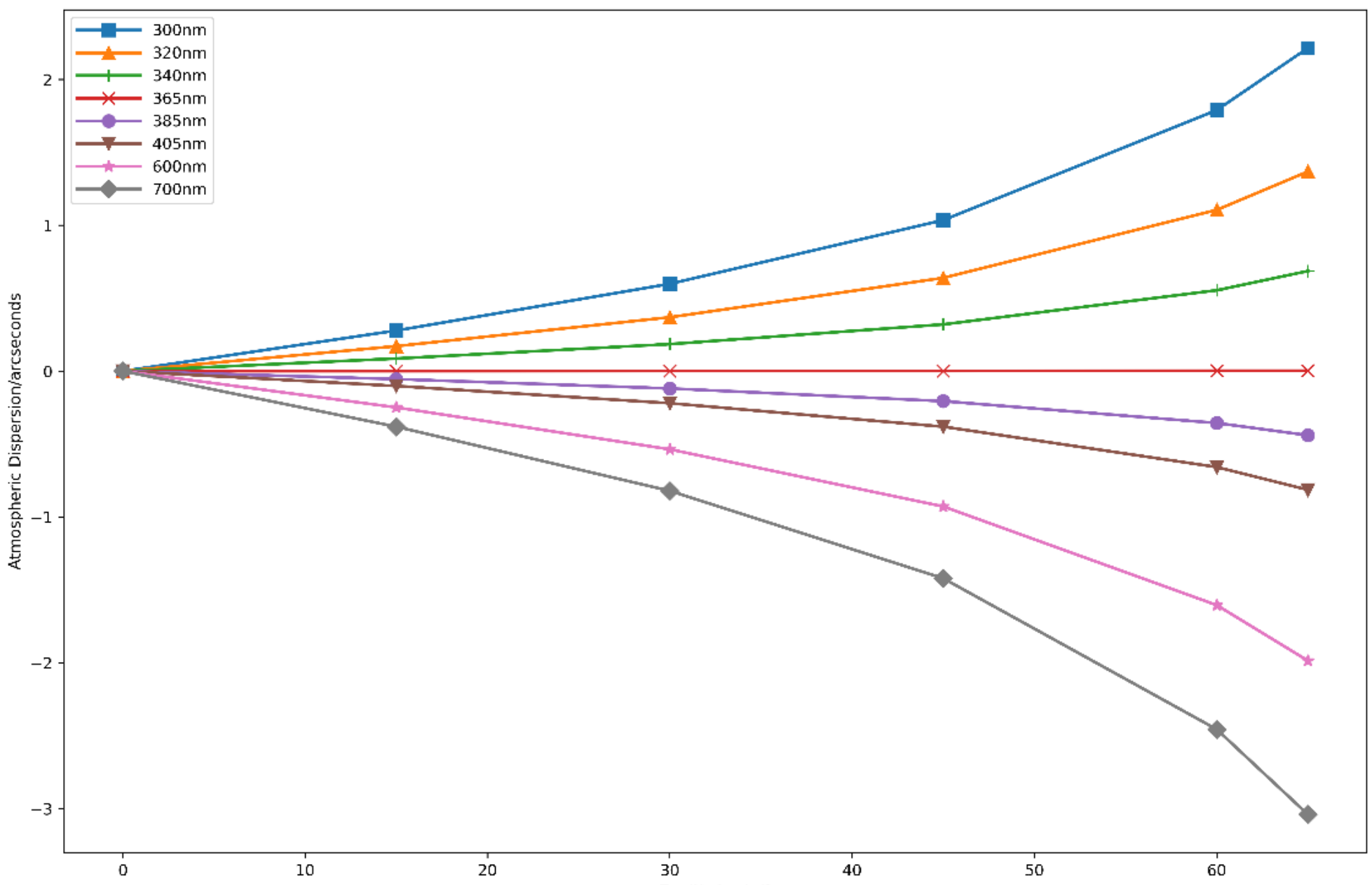

Figure 6. Plot of atmospheric dispersion at ESO Paranal at different zenith angles, modelled using the Ciddor model [4]. 365nm is set as a reference wavelength as it is near the centre of the science wavelength range and matches an emission line produced by the on-board Hg lamp.

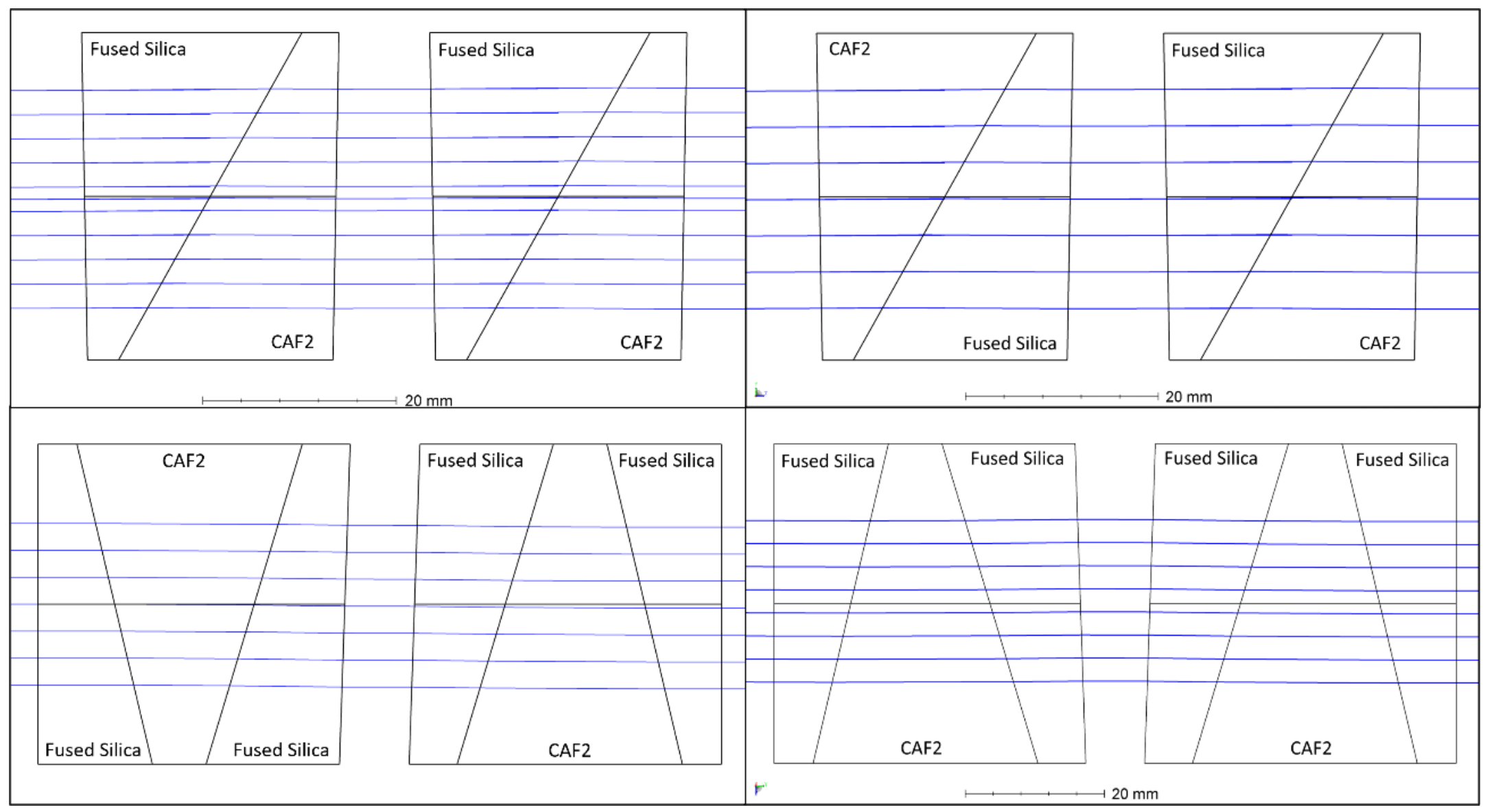

Figure 7. Optical design of the fore-optics and A&G ADCs. (top-left) fore-optics ADC in the non-dispersion configuration – 0° zenith angle, (top-right) fore-optics ADC in the maximum dispersion configuration – 65° zenith angle, (bottom-left) A&G ADC in the non-dispersion configuration – 0° zenith angle, (bottom-right) A&G ADC in the maximum dispersion configuration – 60° zenith angle. CAF2 is shorthand for calcium fluoride.

For the fore-optics ADC, the resulting design is 2 identical fused silica/calcium fluoride amici prisms with the outer faces tilted at 1.03° with respect to the rotation axis, while the angle of the common surface is at 29.16° to the rotation axis. The prism pairs are 34mm in diameter and 26mm thick at their centres. This design has the advantage of creating a non-deviation wavelength (a wavelength which passes undeviated through the ADC for any rotation on-axis) at 365nm. This is useful as it can be used as the guiding wavelength of the telescope and matches an emission line produced by the on-board Hg lamp. The resulting residual dispersion at different zenith angles is shown in Figure 8. The maximum residual of 4.5 milliarcseconds is negligible for a seeing-limited instrument.

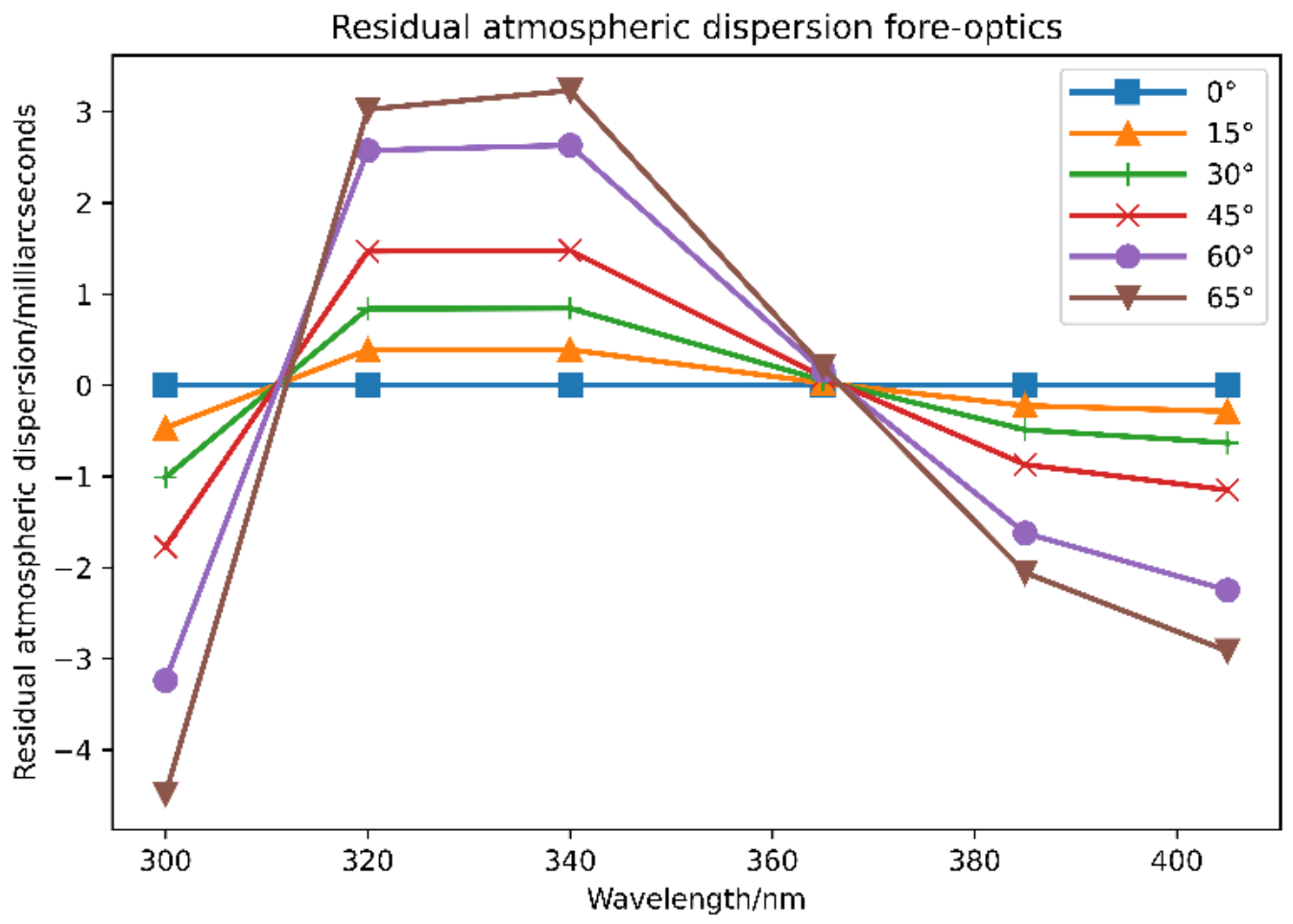

Figure 8. Plot of residual atmospheric dispersion after correction by the ADC for the fore-optics between 300 and 405nm. The non-deviation wavelength is shown at 365nm.

For the A&G ADC, the design is 2 identical double amici prisms composed of a calcium fluoride prism wedged between 2 fused silica prisms. This more complex design is required due to the larger FOV and wavelength range that is required in the A&G sub-system. The first surface is parallel to the rotation axis, while the following surfaces are at 14.03°, -15.90°, and -1.91° to the rotation axis respectively. The prisms are 46mm in diameter and the total centre thickness is 44mm. The resulting residual dispersion at different zenith angles is shown in Figure 9. The design creates two non-deviation wavelengths 365nm and 650nm, this minimises the residuals when guiding in the R-band. The residuals are below the requirement of 150 milliarcseconds ensuring the A&G will be able to acquire and provide secondary guiding for all conditions.

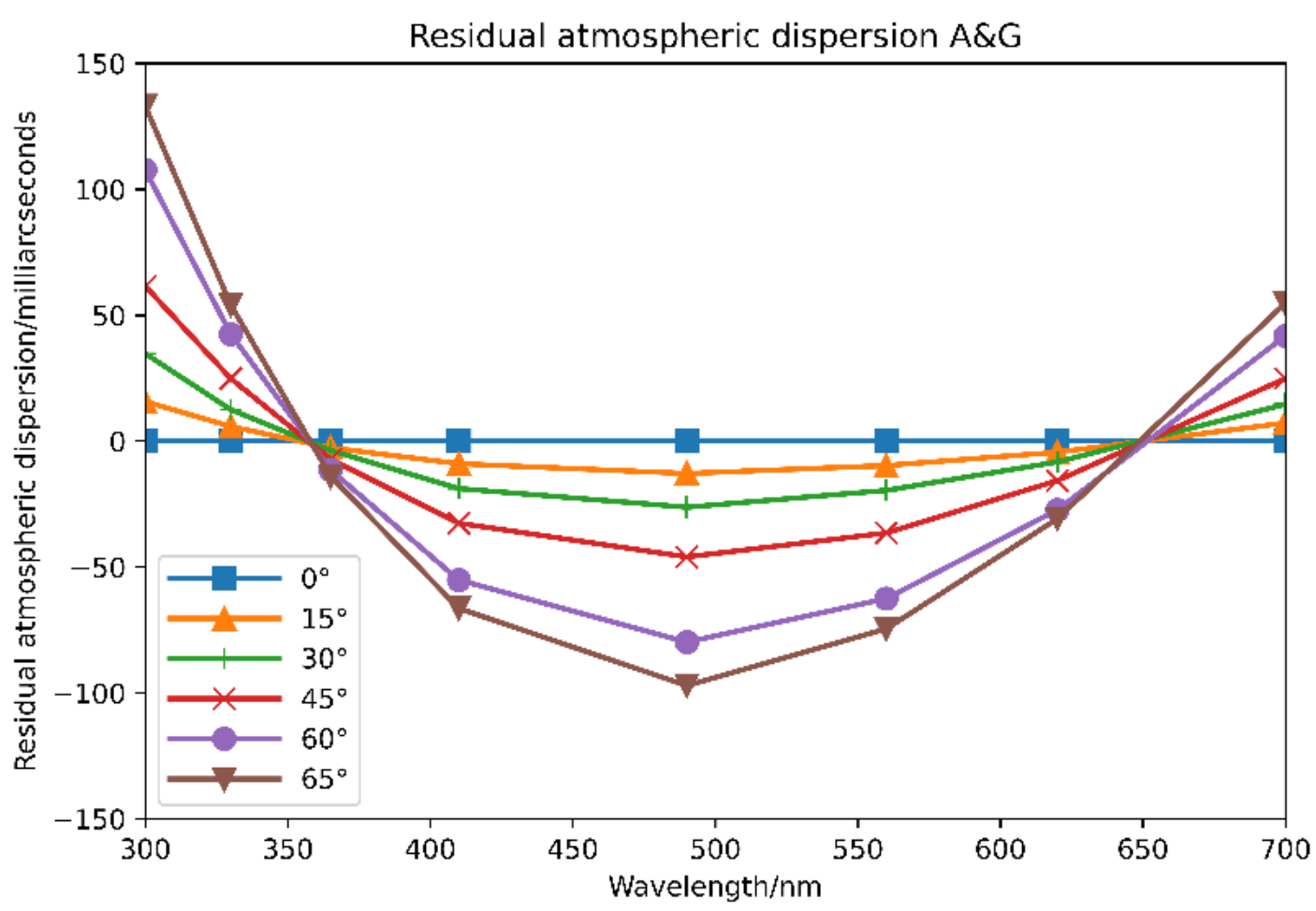

Figure 9. Plot of residual atmospheric dispersion after correction by the ADC for the A&G between 300 and 700nm. The non-deviation wavelengths are shown at 365nm and 650nm.

# 4. IMAGE SLICER

## 4.1 Image Slicer Optical Design

The image slicer re-formats the FOV to generate the entrance slit for the spectrograph. This allows for high spectral resolution to be achieved without the use of adaptive optics in the UV and reduces the sizes of the optical components of the spectrograph. The optical designs of the image slicers are an iteration of the Phase A designs presented in detail in [5]. To achieve the high efficiency required for the instrument, the slicer mirrors will be coated with enhanced aluminium while the camera mirrors will have a multi-layer dielectric coating.

The HR slicer decomposes a FOV of 1.5” x 10”, corresponding to a linear size of 3mm x 20mm at the slicer focal plane, into six slices of 0.25” x 10” (0.5mm x 20mm). The optical design of the HR image slicer is shown in Figure 10.

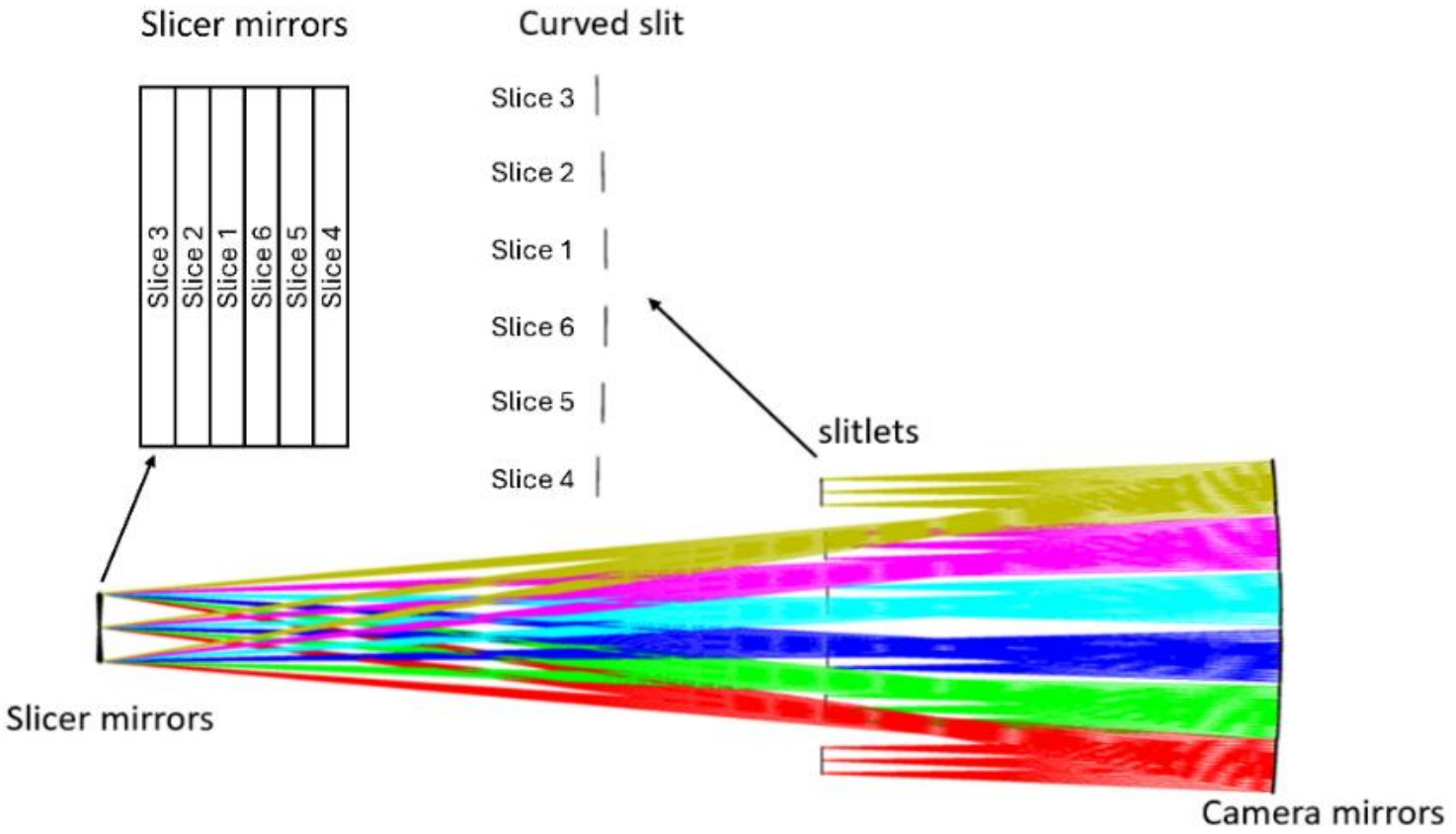

Figure 10. Optical design of HR slicer. The configuration of the slices at the slicer mirror and at the output slit are shown above.

The six slicer mirrors are spherical – each with a concave radius of curvature of 412mm. These feed light onto the six separate spherical camera mirrors (concave radius of curvature 194mm) which focus the light, at a magnification of 0.383, to produce 6 slitlets that form the spectrograph input slit. The slit is ~88mm in length and is slightly curved (maximum difference in z position: 1.8mm) which is corrected for in the spectrograph. The optical performance at the slitlets is shown in Figure 11, showing diffraction-limited performance for the central slices, with acceptable performance throughout, despite some aberrations at the top and bottom of the edge slices. The output beams of the image slicer are designed to overlap on the spectrograph gratings.

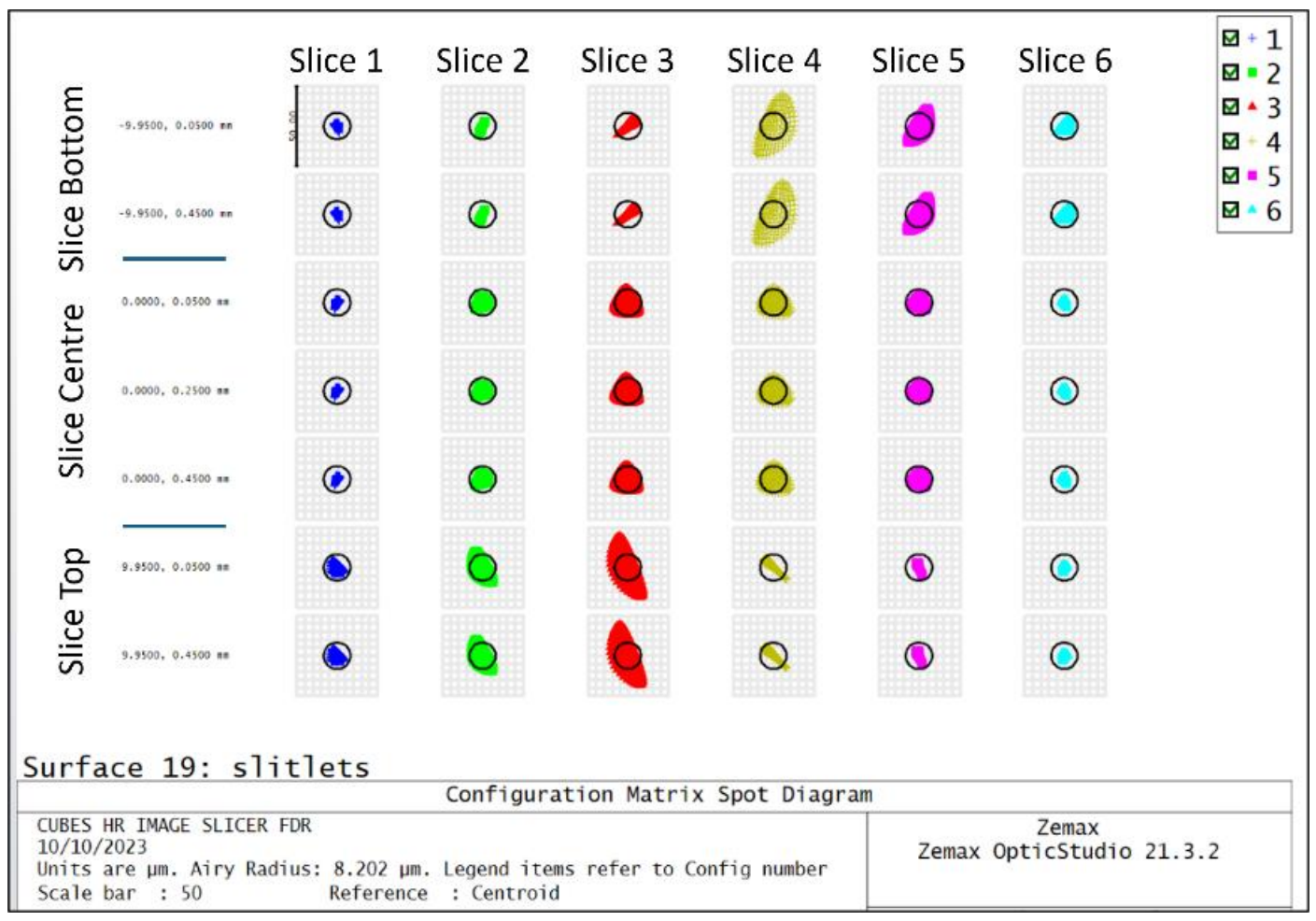

Figure 11. Optical quality of the HR image slicer evaluated at the slitlets position.

The LR slicer decomposes a FOV of 6” x 10”, corresponding to a linear size of 6mm x 20mm at the slicer focal plane, into six slices of 1” x 10” (2mm x 20mm). The optical design mimics that of the HR slicer (same radii of curvature) with the width of the slices 4x larger to accommodate the larger FOV. This produces equivalent optical performance as the HR design but required a re-optimisation of the tilts of the mirrors and a leads to a small lateral decentring (±2mm) of the location of output slitlets.

### 4.2 Active Flexure Compensation (AFC) Projector

The AFC projector takes in Thorium-Argon light coming from an optical fibre from the calibration unit and projects it ~11mm above the output focal plane of each image slicer with an F/# close to 18, where a slit of 0.2 mm x 0.75 mm will be located. This allows for monitoring and compensation of flexures within the spectrograph during exposures.
The projector is fed by a fibre provided by the calibration unit with F/#=3 and core size 300 um. A Thorlabs LA4280 lens (L1) is used to collimate the light from the fibre. A custom plano-convex lens (L2, radius of curvature 22mm) and plano-concave lens (L3, radius of curvature 58mm) then project the image of the fibre onto the slit mask with a magnification of 6.26. The illuminated area on the slit is a circle of 1.8 mm in diameter, to homogenise the slit illumination and to relax the alignment tolerances of the small optics. A pupil mask of 2.5 mm in diameter sets the F/# of the output beam to 18. This F/# is sufficient to over illuminate the grating. A folding mirror (FM) bends the optical beam due to volume constraints.

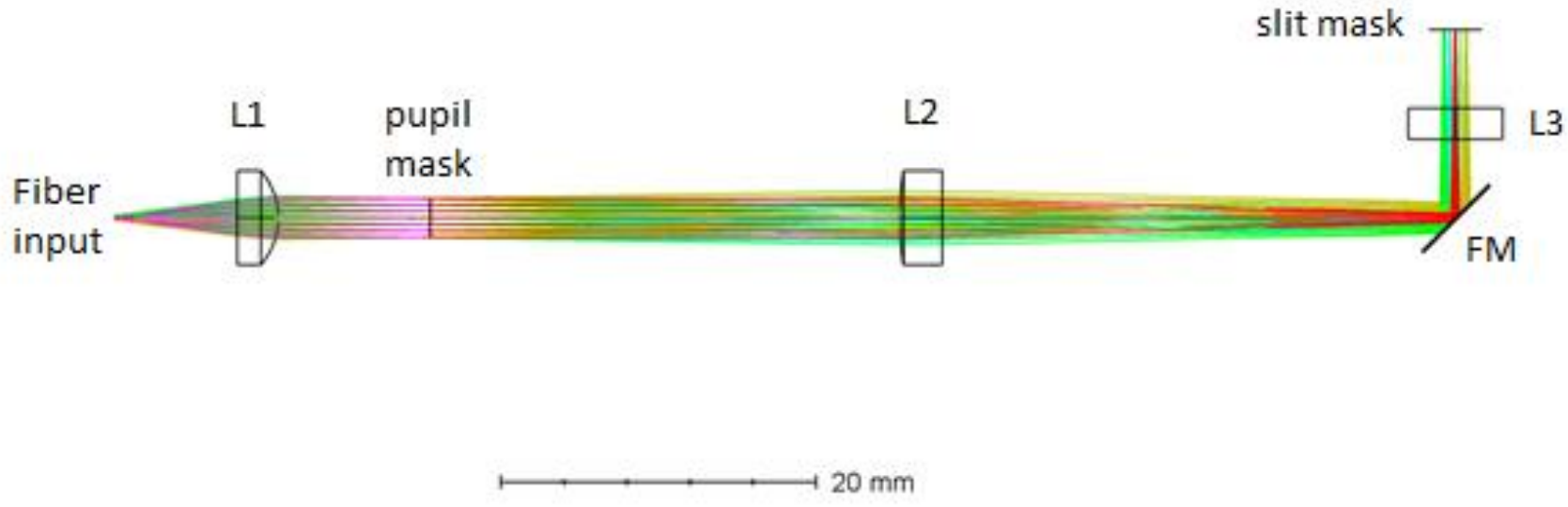

Figure 12: Optical design of the AFC projector

# 5. SPECTROGRAPH

## 5.1 Spectrograph Optics

The spectrograph consists of two arms: a red arm and a blue arm. The light input to the spectrograph is split by a dichroic beamsplitter, which reflects light between 300-351.6nm (blue arm) and transmits light between 345.6-405nm (red arm). The two channels are then steered and collimated before passing through a binary transmission grating with a high line density (3597 lines/mm - blue arm; 3115 lines/mm - red arm) [6]. This diffracts the light in 1st order towards the camera. The dispersed light is then focused by a camera, through a cylindrical detector vessel window, onto a detector.

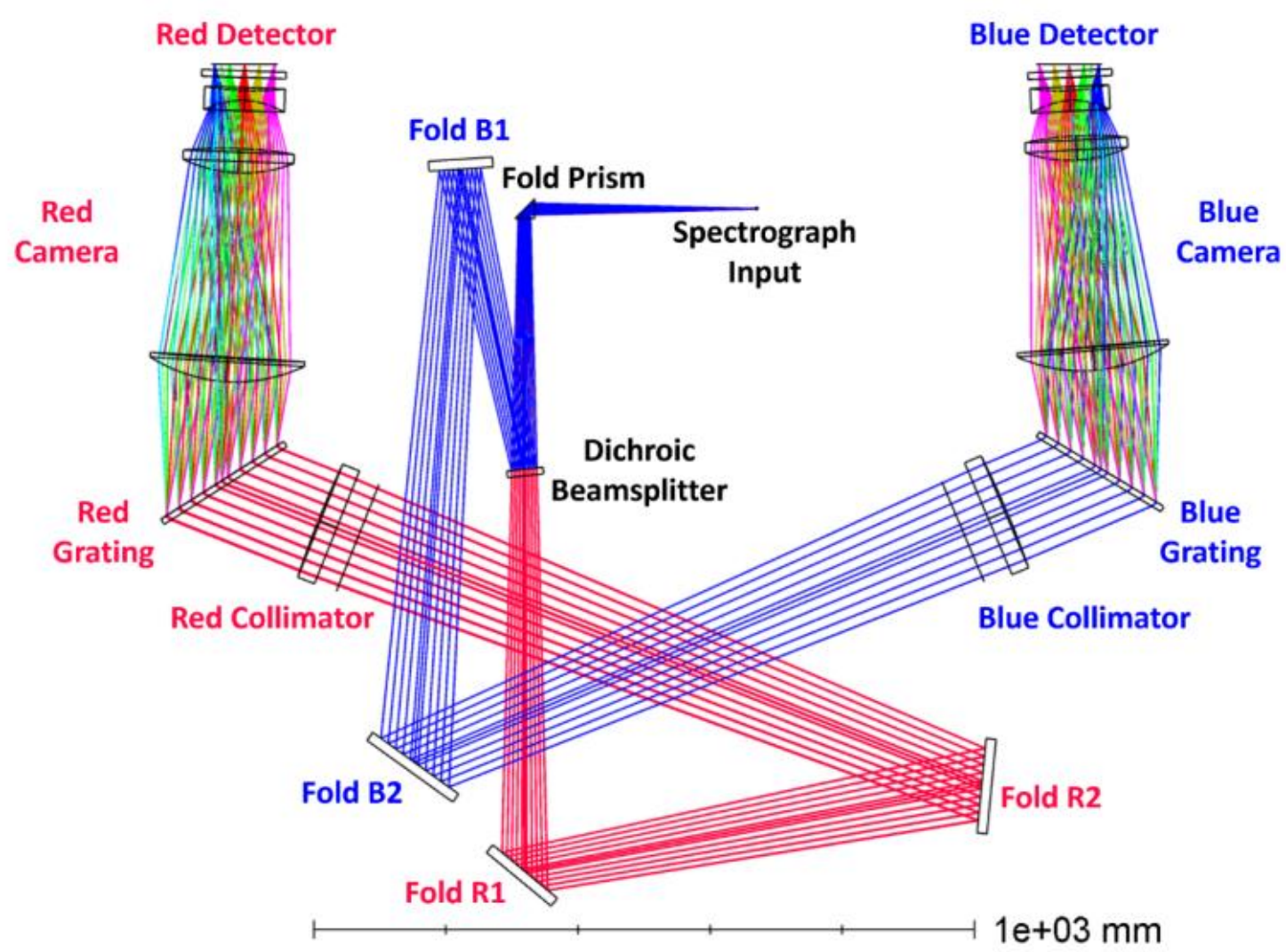

Figure 13. Optical design of the CUBES spectrograph, showing both red and blue arms. The rays are coloured by configuration up to the grating for clarity, then coloured by wavelength after to illustrate the dispersion from the grating.

## 5.2 Spectrograph Camera Design

The spectrograph cameras are designed to reimage the output slit of the image slicer onto the detector with a magnification of 0.12 at the blue end of each arm, to provide 2.3 pixels spectral sampling at $\lambda_{min}$. The CUBES cameras are each composed of 3 fused silica lenses mounted in an aluminium barrel. To achieve the required image quality each lens is composed of one spherical and one conic surface (with modest conic constants between -0.416 and -1.862). Inside the barrel, lenses 2 and 3 are offset a small amount laterally, and angularly, in the dispersion direction with respect to the first lens to provide the required aberration correction. A cylindrical cryostat window is positioned 20 mm after the final lens in the camera barrel. This corrects for residual astigmatism in the camera lens so that required image quality can be achieved. The detector is positioned 10mm from the centre of the cryostat window. It is tilted with respect to the window to account for the wavelength dependent focal length of the camera. After a manufacturing design review, thermal compensators have been included in the camera barrel design to improve the spectral format stability due to thermal variations in focal length. The resulting spectrum has a length of 92.2mm for each arm, ensuring it fits on the 9232 10-micron detector pixels. The height across all six slices, including gaps between slices, is 12.84mm.

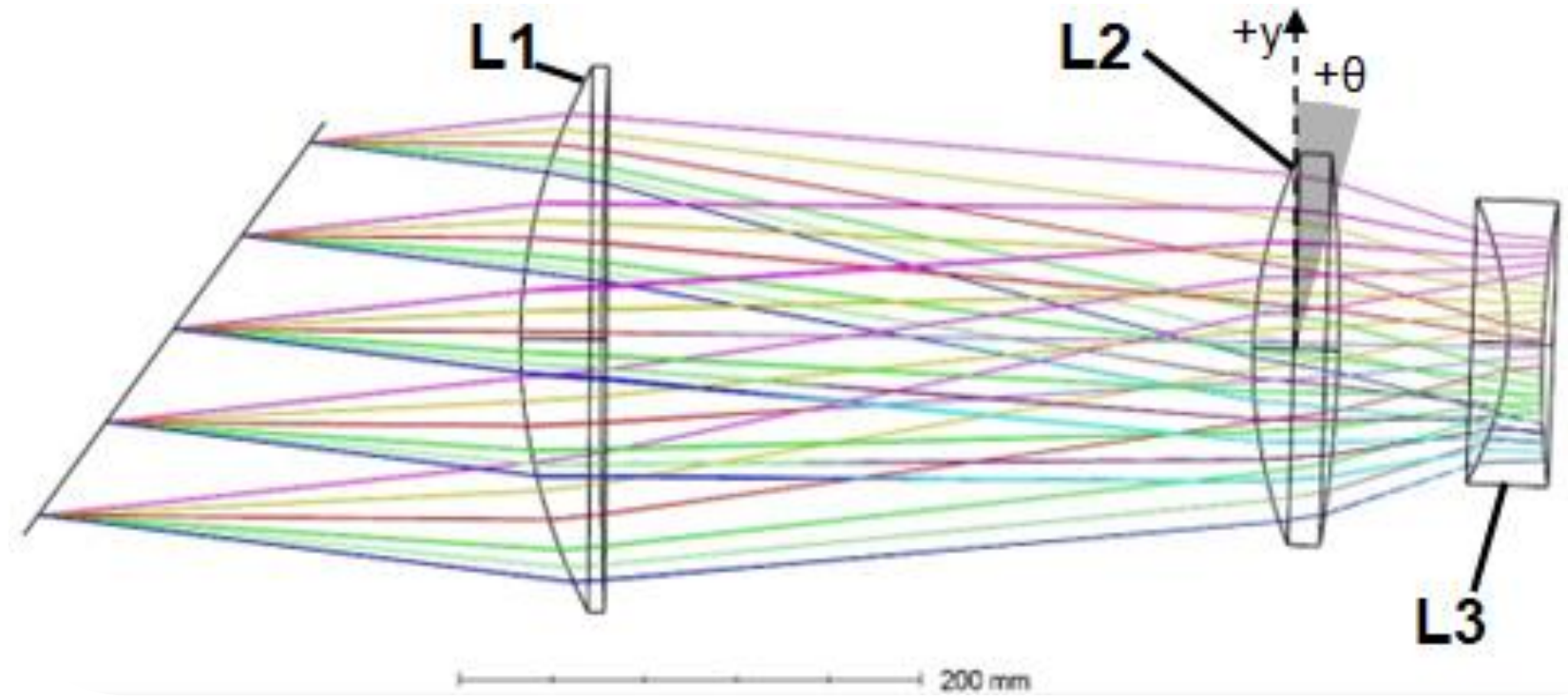

Figure 14. Optical layout of the blue CUBES spectrograph camera, showing the tilt and decentre of lenses L2 and L3 that is needed to correct for chromatic aberrations in the camera.

The optical performance of the blue and red arms of the spectrograph is shown in Figure 15 and Figure 16.

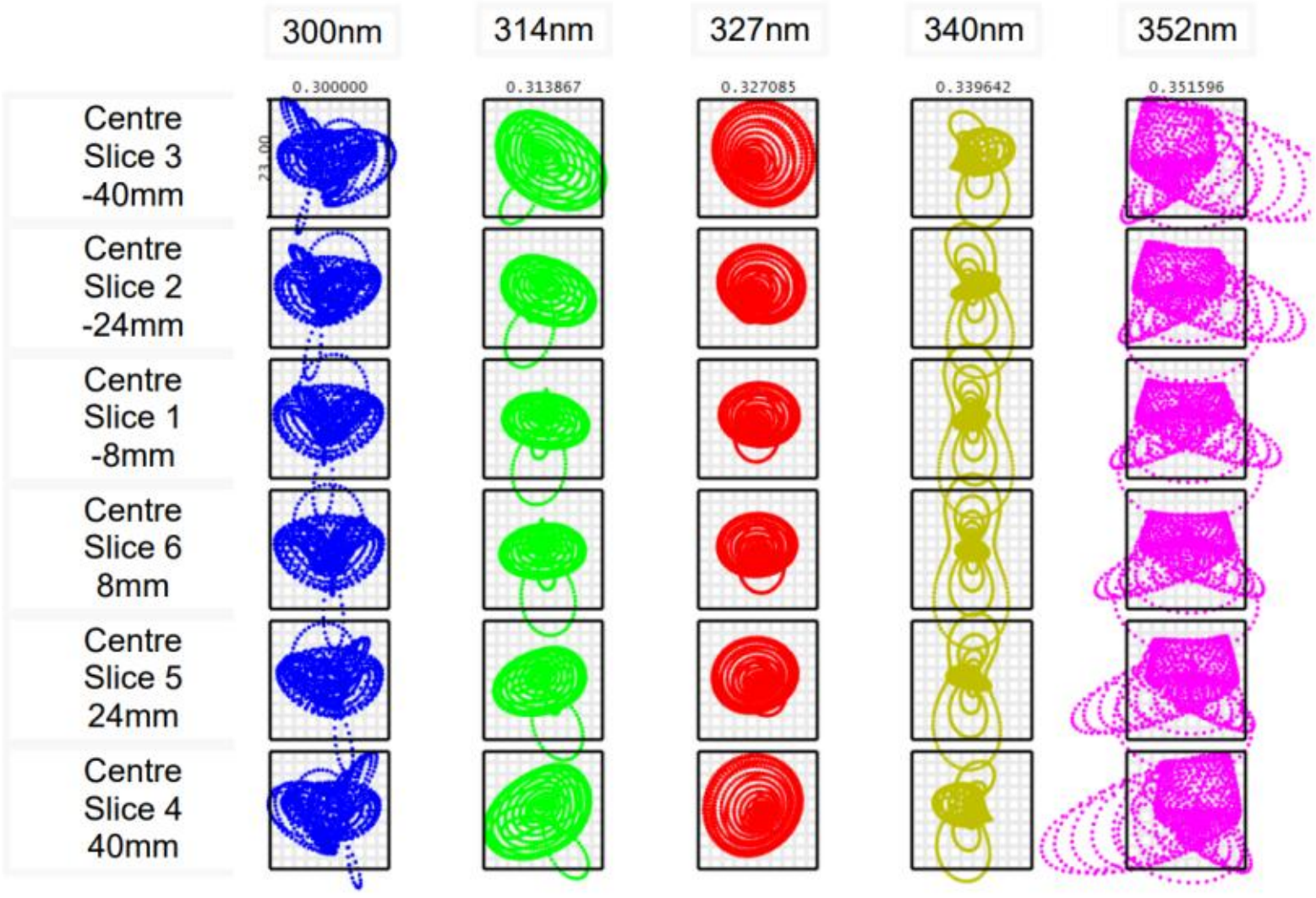

Figure 15. Spot diagram of the blue arm, for different wavelengths at the centre of each slice. Box size 23µm (2.3 pixels)

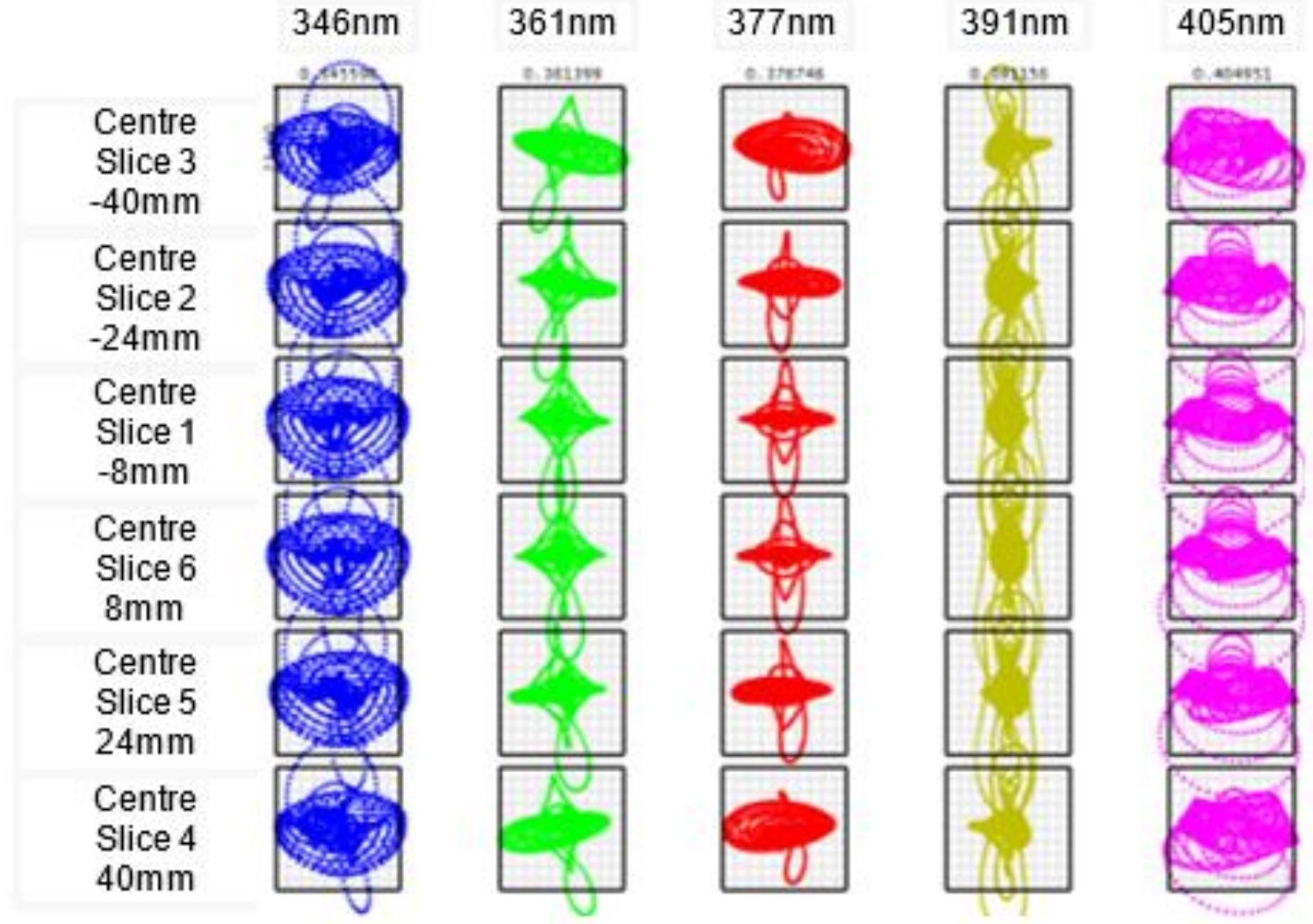

Figure 16. Spot diagram of the red arm, for different wavelengths at the centre of each slice. Box size 23µm (2.3 pixels)

Modelling of the line-spread function including the geometric slit width, detector diffusion, and flexure effects allows the definition of an optical budget for the spectral FWHM of the geometric PSF (modelled as the geometric enslitted energy in Zemax). This can then be compared to the modelled performance for the blue and red arms as shown in Table 1. This shows that the performance exceeds R=20,000 across the entire wavelength range leaving margin for manufacturing and alignment tolerances. The resolving power of the spectrograph design is verified in Genoni et al. [7].

Table 1. Table of the required spectral FWHM of the geometric PSF to reach a resolving power of 20,000 and the as-designed spectral FWHM of the geometric PSF at the centre of each slice. The values given are in microns.

| | | **Wavelength/nm** | | | | | | | | | |
|---|---|---|---|---|---|---|---|---|---|---|---|
| | | **Blue Arm** | | | | | **Red Arm** | | | | |
| | | **300** | **314** | **327** | **340** | **352** | **346** | **361** | **377** | **391** | **405** |
| **Requirement** | | 12.4 | 15.9 | 19.1 | 22.6 | 26.4 | 13.1 | 16.4 | 19.6 | 22.4 | 26.0 |
| **Slice** | **3** | 9.9 | 8.0 | 9.6 | 5.0 | 15.4 | 9.4 | 5.9 | 7.4 | 6.4 | 11.9 |
| | **2** | 10.6 | 5.8 | 7.3 | 6.2 | 12.4 | 11.0 | 5.3 | 5.3 | 8.9 | 9.1 |
| | **1** | 11.1 | 5.3 | 6.5 | 7.4 | 11.2 | 11.7 | 5.4 | 4.4 | 10.2 | 8.0 |
| | **6** | 10.5 | 5.3 | 6.8 | 7.0 | 11.5 | 11.3 | 5.1 | 4.6 | 9.9 | 8.2 |
| | **5** | 9.5 | 7.2 | 8.8 | 4.6 | 13.4 | 9.6 | 5.3 | 6.3 | 7.5 | 9.8 |
| | **4** | 10.8 | 11.1 | 12.5 | 5.7 | 17.5 | 9.6 | 7.5 | 9.4 | 4.9 | 13.3 |

## 6. CONCLUSIONS

This proceeding presented the status of the optical design of the Cassegrain U-Band Efficient Spectrograph (CUBES) for the ESO VLT after the final design review. The fore-optics and A&G sub-system designs were presented and the performance of the dispersion correction and image quality were demonstrated to meet the requirements of the instrument. The optical designs of the image slicers and the AFC projector were summarized, and the spectrograph optical design and an analysis of the optical performance were described. The optical design is now mature and ready for the upcoming MAIT (Manufacturing, Assembly, Integration, and Test) phase.